\documentclass[prl,twocolumn,tightenlines,superscriptaddress,preprintnumbers,floatfix,nofootinbib,]{revtex4}
\usepackage{graphicx}
\usepackage{amsmath}
\usepackage{amssymb}
\usepackage{bbold}
\usepackage{color}
\input epsf

\def\p{\mathbf{p}}
\def\k{\mathbf{k}}
\def\pp{\mathbf{p'}}
\def\kp{\mathbf{k'}}

\newcommand{\be}{\begin{equation}}
\newcommand{\ee}{\end{equation}}
\newcommand{\h}{\mathbf{h}}
\newcommand{\q}{\mathbf{q}}
\begin{document}
\newcommand{\nn}{\nonumber \\}
\renewcommand{\Im}{{\rm Im}}
\renewcommand{\Re}{{\rm Re}}
\newcommand{\Li}{{\rm Li}}
\newcommand{\6}{\partial}

\title{Isotropization and  hydrodynamization in weakly coupled heavy-ion collisions}
\preprint{CERN-PH-TH-2015-142}
\author{Aleksi Kurkela}
\affiliation{Physics Department, Theory Unit, CERN, CH-1211 Gen\`eve 23, Switzerland}
\affiliation{Faculty of Science and Technology, University of Stavanger, 4036 Stavanger, Norway}
\author{Yan Zhu}
\affiliation{Departamento de Fisica de Particulas and IGFAE, Universidade de Santiago de Compostela, E-15706 Santiago de Compostela, Galicia, Spain}
\begin{abstract}
We numerically solve 2+1D effective kinetic theory of weak coupling QCD under longitudinal expansion relevant for early stages of heavy-ion collisions. We find agreement with viscous hydrodynamics and classical Yang-Mills simulations in the regimes where they are applicable. By choosing  initial conditions that are motivated by color-glass-condensate framework we find that for $Q_s=2$GeV and $\alpha_s=0.3$, the system is approximately described by viscous hydrodynamics well before $\tau \lesssim 1.0$ fm/c. 
\end{abstract}
\maketitle
\section{Introduction}
In the weak coupling picture of the pre-thermal evolution
of heavy-ion collisions, the post collision debris that
end up in the mid-rapidity region undergo several stages
that are characterised by widely different physics and degrees 
of freedom. 
On the one hand, according to the saturation paradigm \cite{CGC}, at very early
times $\tau \sim Q_s ^{-1}$, where $Q_s$ is the typical energy scale right after the collision, the energy is deposited in strong
color fields.  These strong fields admit a description
in terms of classical Yang-Mills theory to leading order in 't Hooft coupling $\lambda=4\pi N_c \alpha_s$. Indeed, there have 
been several interesting studies of classical Yang-Mills fields under longitudinal 
expansion in the past years \cite{Epelbaum:2013waa,Berges:2013eia, Berges:2013fga, Gelis:2013rba,Berges:2014yta}. 
On the other hand, once the system has reached a local thermal equilibrium,
or at least approximately isotropized \cite{Arnold:2004ti}, the matter in the mid-rapidity
region (at sufficiently low $p_T$) is described by relativistic fluid dynamics \cite{oldhydro,newhydro}.   
There has been a sizeable and very successful program of numerical simulations
of relativistic hydrodynamics.

It is well known that classical Yang-Mills theory can not reach thermalization due 
to the Rayleigh-Jeans catastrophe nor can it isotropize when exposed to rapid longitudinal expansion as noticed by \cite{Baier:2000sb}. Instead classical evolution drives
the system further away from equilibrium, making the system less occupied (i.e. weaker fields) but more anisotropic, which has also been observed in the simulations of  \cite{Berges:2013fga,Berges:2013eia}.
Therefore, there is a missing link between the physics of saturated gluon fields and fluid dynamics
that needs to be bridged in order to fully carry the predictions of saturation physics to the hydrodynamic regime. This gap can be filled with the systematically improvable framework of effective kinetic theory (EKT) of \cite{Arnold:2002zm}.

The EKT faithfully describes to leading order in $\lambda f$ systems where the typical occupancies of gluons are not nonperturbative $f\ll 1/\lambda$ and have momentum significantly larger than the in-medium screening scale $p^2>m^2 \equiv \lambda \int_{\mathbf{p}} f(p)/p$. At weak coupling, these conditions are certainly fulfilled in thermal equilibrium and therefore we can describe the system all the way to the equilibrium. While these conditions are not fulfilled at the very earliest times, both EKT and classical Yang-Mills give an equally valid leading order description for a wide range of large but perturbative occupancies $1\ll f \ll 1/\lambda$ \cite{Mueller:2002gd,York:2014wja}.
Therefore a possible strategy to simulate the system through all time scales is to start the simulation with a classical Yang-Mills simulation (for $Q_s\tau \sim1 $ and $f\sim 1/\lambda$) and subsequently pass the system to EKT at some arbitrary time $ \tau_{EKT} Q_s \gg 1$. Then, once EKT has brought the system sufficiently close to the thermal equilibrium, both hydrodynamics and EKT should give equivalent descriptions and the system can be passed to a hydrodynamical simulation at some arbitrary time $\tau_{hydro}$. 

The proof of principle of such a procedure was shown in a series of papers \cite{Kurkela:2012hp,York:2014wja,Kurkela:2014tea} in an isotropic setting in the absence of expansion. Here, we present first results of numerical simulations of the EKT in a boost invariant 2+1D setting and demonstrate the connection with both classical Yang-Mills simulations and viscous hydrodynamics. We make a first attempt to model the early stages of heavy-ion collision starting from the over-occupied region all the way to the thermal equilibrium and find that after time $Q_s\tau\sim 5.0$ the time evolution of energy density is described by viscous hydrodynamics to a better than $10\%$ accuracy for a realistic coupling $\lambda =10 $ corresponding to $\alpha_s \approx 0.3$.

The current result is, however, incomplete in two ways. Firstly, there have been significant steps to describe the melting of the strong color-fields in classical statistical field theory \cite{Gelis:2013rba,Epelbaum:2013waa,Berges:2014yta}, but currently we do not have a reliable initial state from a 3+1D simulation to enter into our EKT simulation.  Therefore we initialize our system at $Q_s \tau\sim 1$ with the energy density given by 2+1D simulation \cite{Lappi:2011ju} and vary the parameters of the initial condition to quantify the ignorance of the very early time dynamics. Secondly, certain non-perturbative chromo-Weibel instabilities \cite{Mrowczynski:1988dz} arising from anisotropic screening play a parametrically leading order role during the whole non-equilibrium evolution \cite{Kurkela:2011ti,Kurkela:2011ub}. However, numerical simulations of classical fields \cite{Berges:2014yta} have not been able to see them beyond the very early times, suggesting that even if parametrically leading order, their effect is numerically
small for values of $\lambda$ that are phenomenologically interesting.  Therefore, as a formally correct treatment of anisotropic screening is rather complicated \cite{Romatschke:2003ms,Romatschke:2006bb,Arnold:2005vb,Bodeker:2007fw} and not a fully solved problem, we in this first work will treat the screening in an isotropic way. Therefore, our results are correct to leading order in $\lambda$ for systems that are close to isotropy (late times) and for large anisotropies our results have leading logarithmic accuracy apart from the instabilities.

\section{Methodology}
The EKT of \cite{Arnold:2002zm} is defined though the effective Boltzmann equation for the color and spin averaged distribution function of gluons, and to leading order in $\lambda$ it contains effective $2\leftrightarrow 2$ scattering and $1 \leftrightarrow 2 $ splitting terms. 
\begin{align}
-\frac{d f_\p}{d\tau} = \mathcal{C}_{1\leftrightarrow 2}[f_\p] + \mathcal{C}_{2\leftrightarrow 2}[f_\p] + \mathcal{C}_{\rm exp }[f_\p].
\end{align}
We restrict ourselves to azimuthally symmetric distributions but allow anisotropy in the $\mathbf{z}$-direction so that it is enough to specify $f_\p=f_{x_p, p}$ with $x_p \equiv {\hat{\bf z}}\cdot \hat{\p}$. The effect of longitudinal expansion is encapsulated in  $\mathcal{C}_{\rm exp}[f](\p) = - \frac{p_z}{\tau}\frac{\partial }{\partial p_z} f(\p)$ \cite{Mueller:1999pi}.

The $2\leftrightarrow 2$ effective scattering term reads
\begin{align}
 \mathcal{C}_{2\leftrightarrow 2}&[f](\tilde{\p})=  \frac{(2\pi)^3}{ 4\pi \tilde p^2 }  \frac{1}{8 \nu}\int d\Gamma_{PS}|\mathcal{M}|^2 \nonumber
 \\
&  \times \left(f_\p \, f_\k \, g_\pp \, g_\kp - f_\pp \, f_\kp \, g_\p \, g_\k \right)\label{C22}
 \\
&\times \left[ \delta(\tilde{\p}-\p)+\delta(\tilde{\p}-\k)-\delta(\tilde{\p}-\pp)-\delta(\tilde{\p}-\kp)\right], \nonumber
\end{align}
where $\nu = 2 d_A$ and $g_\p \equiv 1+f_\p$. $d \Gamma_{PS}$ is the integral measure over the phase space of $2\leftrightarrow 2$ processes, singling out the $z$-direction
\begin{align}
\int d\Gamma_{PS} \equiv \frac{1}{2^{11}\pi^7}&\int_0^\infty dq \int_{-q}^{q}d\omega \int_{(q-\omega)/2}^{\infty}
dp \int_{(q+\omega)/2}^{\infty}dk \nonumber \\ 
&\times \int_{-1}^{1} d x_q \int_{0}^{2\pi}d \phi_{pq}d \phi_{kq}. \label{gamma_PS}
 \end{align}
 In these coordinates the angles of incoming and outgoing momenta read
\begin{equation}
x_{\{p\}}= -\sin\theta_{\{p\}q}\cos\phi_{\{p\}q} \sqrt{1-x_q^2}+\cos\theta_{\{p\}q}x_q\;.
\end{equation}
This relation holds for $x_k$, $x_{p'}$, and $x_{k'}$ with $\cos \phi_{p'q} = \cos \phi_{pq}$,  $\cos \phi_{k'q} =  \cos \phi_{kq}$,
\begin{align}
\cos\theta_{pq} = \frac{\omega}{q} + \frac{t}{2 pq}, & \quad 
\cos\theta_{kq} = \frac{\omega}{q} - \frac{t}{2 kq},\\
\cos\theta_{p'q} = \frac{\omega}{q} - \frac{t}{2 p' q}, & \quad
\cos\theta_{k'q} = \frac{\omega}{q} + \frac{t}{2 k' q},
\end{align}
and $t \equiv \omega^2 - q^2$. The effective matrix element for most kinematics is the ordinary vacuum tree-level element
\begin{align}
\frac{|\mathcal{M}|^2}{16 \lambda^2 d_A} =  \left( \frac{9}{4} + \frac{(s-t)^2}{u^2}+ \frac{(u-s)^2}{t^2}+ \frac{(t-u)^2}{s^2}\right).
\end{align}
For small $t \sim m^2$ or $u \sim m^2$ the tree-level element has an IR divergence that is regulated by the physics of screening. In the formulation of \cite{Arnold:2002zm}, the screening is treated by replacing the naive tree-level matrix element by the retarded Hard Thermal Loop (HTL) resummed expression in the soft kinematic region.
The resummed propagator in the anisotropic case, however, contains 
a non-integrable singularity, signalling the presence of an instability in the system. Therefore, in the anisotropic system this procedure is not fully satisfactory. Here, in this work we will restrict ourselves 
to isotropic screening and use the prescription derived in \cite{York:2014wja}, replacing in the denominator (similarly for $u$)
\begin{align}
q^2 t \rightarrow t(q^2 + 2 \xi_0^2 m^2),
\end{align} 
with $\xi_0 = e^{5/6}/\sqrt{8}$, which is leading order accurate in the case of isotropic distribution.
 
The effective $1\leftrightarrow 2$ collinear splitting term reads
\begin{align}
&\mathcal{C}_{1\leftrightarrow 2}[f](\tilde{\p}) = 
\frac{(2\pi)^3}{ 4\pi \tilde p^2 } \frac{1}{\nu}\int_0^{\infty}dp\int_0^{p/2}dk' \left[ 4\pi \gamma(p; p', k')\right]\nonumber \\
&\times \left( f_{x_p,p} g_{x_p,p'} g_{x_p,k'} - g_{x_p,p} f_{x_p,p'} f_{x_p,k'} \right) \nonumber \\
& \times  \left[ \delta(\tilde{p}-p)-\delta(\tilde{p}-p')-\delta(\tilde{p}-k')\right],
\label{C12}
 \end{align}
where $p'=p-k'$, and the effective splitting rate reads
\begin{align}
\gamma(p; p',k') = \frac{p^4 + p'^4+k'^4}{p^3 p'^3 k'^3 }\frac{d_A \lambda}{2 (2\pi)^3}\int \frac{d^2 h}{(2\pi)^2} 2 {\bf h} \cdot {\rm Re} {\bf F}.\nonumber
\end{align} 
Again, consistently assuming only isotropic screening, the function ${\bf F}$ is given by the solution to the linear integral equation
\begin{align}
&2 {\bf h} =  i \delta E({\bf h} ){\bf F}({\bf h})+\frac{\lambda T_*}{2}\int \frac{d^2  q}{(2\pi)^2} \Big[\mathcal{A}({\bf q}) \\
&\times \left( 3{\bf F({\h})}-{\bf F}({\h}-p \q)-{\bf F}({\h}-k \q)-{\bf F}({\h+p' \q})\right)\Big]\nonumber.
\end{align}
$
T_* = \frac{\lambda}{2 m^2}\int \frac{d^3 p}{(2\pi)^3}f_\p(1+f_\p) ,
$
and $\delta E = m^2/p'+m^2/k'-m^2/p + {\bf h}^2/2p k' p'$.
We solve this equation using an efficient numerical method introduced in \cite{Ghiglieri:2014kma}. 

For the numerical solution of the EKT, we discretize $n_{x_p,p}= 4\pi p^2 /(2\pi)^3 f_{x_p,p}$ on a 2D grid and Monte Carlo estimate the integrals of Eqs.~(\ref{gamma_PS},\ref{C12}). We use a logarithmically spaced grid for both $p$ and $x_p$ with at least 250 grid points in angular direction and at least 100 in the $p$-direction. We have varied the number of grid points to verify that the results are insensitive to the number of grid points. 
Our algorithm, the discrete-$p$ method of \cite{York:2014wja}, conserves energy exactly and has exactly correct particle number violation for $1\leftrightarrow 2$ term.

\begin{figure}
\includegraphics[width=0.45\textwidth]{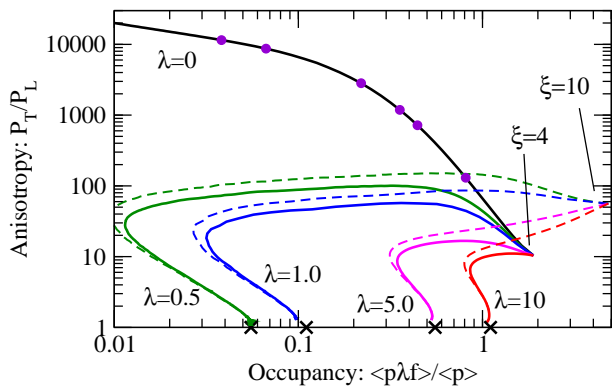}
\caption{Trajectories of runs with different initial conditions $\xi=4$ (Solid lines) and $\xi=10$ (dahsed lines) and varying coupling $\lambda$ in a plane of mean occupancy (weighted by the energy of particles) and anisotropy. The $\lambda=0$ line corresponds to classical field approximation. The violet dots refer to the times in Fig.~\ref{pz_scaling}. The simulations at finite coupling reach thermal equilibrium located at points indicated by the black crosses.}
\label{trajectories}
\end{figure}

\section{Results}
We shall now apply EKT to simulate the prethermal evolution of the expanding
fireball created in a heavy-ion collision. In saturation framework, the initial condition
is typically described in terms of ``gluon liberation coefficient'' $c$ and mean transverse
momentum $\langle p_T \rangle/Q_s$ \cite{Mueller:1999fp,Kovchegov:2000hz}.
 The gluon liberation coefficient is proportional 
to the total gluon multiplicity per unit rapidity 
\begin{align}
  2 d_A \tau \int \frac{d^3p}{(2\pi)^3} f  \equiv \frac{d N_{init. g}}{d^2 {\bf x}_\perp d y} = c \frac{d_A Q_s^2 } {\pi \lambda}   \label{eq:c}
\end{align}
after the classical fields have decohered and can be described
in terms of quasi-particles. 
Lappi \cite{Lappi:2011ju} finds in JIMWLK evolved MV model values relevant for heavy-ion collisions relevant for LHC of roughly $\langle p_T \rangle \approx 1.8 Q_s$ and $c \approx 1.25$ extracted at time $Q_s\tau = 12$ from a 2D classical Yang-Mills simulation.
By construction the distribution then has $\langle p_z \rangle = 0$. But it is has been noted \cite{Romatschke:2005pm} that 
certain plasma instabilities will broaden the distribution in $p_z$ in a time scale $Q \tau \sim 1/\log^2(\lambda^{-1})$. Therefore, as a rough estimate of the initial condition we instead take somewhat arbitrarily our initial condition at the time $Q\tau = 1$ to be
\begin{align}
f(p_z,p_t) &= \frac{2}{\lambda} A f_0(p_z \xi/\langle p_T \rangle, p_\perp/\langle p_T\rangle ),\\
f_0(\hat p_z,\hat p_\perp) & = \frac{1}{\sqrt{\hat p_\perp^2+\hat p_z^2}} e^{-2 (\hat  p_\perp^2 + \hat  p_z^2)/3}, \label{f0}
\end{align}
 choosing $A$ such that comoving energy density $\tau \epsilon = \langle p_T \rangle dN/d^2{\bf x} dy$ is fixed. We then vary $\xi=4,10$ to quantify our ignorance of the initial nonperturbative dynamics. 

\begin{figure}
\includegraphics[width=0.35\textwidth]{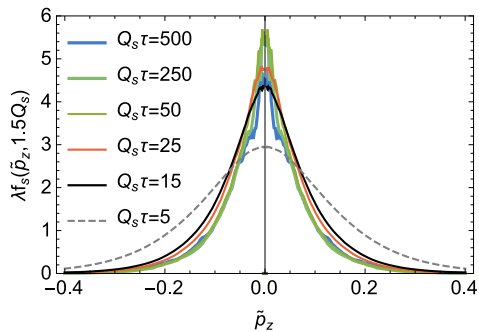}
\caption{
Sections of scaled distribution $f_s(\tilde{p}_z,p_\perp) = (Q_s \tau)^{2/3}f(\tilde{p}_z \,(Q_s\tau)^{-1/3}, p_\perp)$ at $p_\perp = 1.5 Q_s$ in classical approximation at vastly different times. The good overlap of the curves indicates that system has reached the classical scaling solution of Eq.~(\ref{eq:scaling}). In contrast, $Q_s \tau=5$ has not yet reached the scaling solution.
\label{pz_scaling}
}
\end{figure}

Figure \ref{trajectories} displays a set of trajectories from simulations with varying $\lambda$
and $\xi =4, 10 $ on a plane of mean occupancy (weighted by the energy of particles) and anisotropy measured by the ratio of the transverse and longitudinal pressures $P_T/P_L$. 
The line with $\lambda=0$ corresponds to the classical field limit  $\lambda \rightarrow 0 $ with fixed $\lambda f$, which is obtained in EKT by including only the highest power of $f$'s in Eqs.(\ref{C22},\ref{C12}). 
The classical field theory can not thermalize 
and indeed instead
it flows to a stationary scaling solution.  By performing classical Yang-Mills simulations Berges et. al have established that the scaling solution can described by a scaling form of the distribution function \cite{Berges:2013fga},
\begin{align}
f(p_z,p_\perp,\tau)= (Q_s\tau)^{-2/3} f_S((Q_s \tau )^{1/3}p_z,p_\perp),
\label{eq:scaling}
\end{align}
where $f_S$ is approximately constant as a function of time. 
This behavior is demonstrated
in Figure \ref{pz_scaling} where we plot a section of rescaled distribution function $f_S$ measured at various times as a function for $\tilde{p}_z \equiv (Q_s\tau)^{1/3}p_z$ at fixed $p_\perp$ following Berges et al.. Our results corroborate that such a scaling solution exists at late times within the classical approximation and we observe that the scaling regime is reached after a time $Q_s\tau \sim 15 $.

Moving on to the finite but small coupling $\lambda=1,0.5$, we see qualitative agreement with the parametric picture of bottom-up thermalization of \cite{Baier:2000sb}: there are three distinct stages of evolution visible. In the first stage the classical evolution drives the system more anisotrpic and less occupied. Once the occupancies reach $f\sim 1$, there is a qualitative change in the dynamics of the system as Bose enhancement is lost. This has the effect that ansiotropy freezes but the system still continues to get more dilute. Only in the last stage which is characterized by a radiational break up of the particles at the scale $Q_s$, does the trajectory turn back and reach thermal equilibrium, denoted by the black crosses in the Figure \ref{trajectories}.  For larger values of coupling $\lambda = 5.0,10$, these features become however less pronounced and the system takes more straight trajectory towards equilibrium. It should be noted that for these values of $\lambda$ the assumption of $p \gg m$ is not satisfied and large NLO corrections are to be expected. 

\subsubsection{Approach to hydrodynamics}

We expect that 
late time 
evolution should be described by relativistic hydrodynamics. Under flow with translational invariance along transverse directions and boost invariance, the hydrodynamical relations read to second order in gradients \cite{Baier:2007ix}
\begin{align}
\partial_\tau \epsilon &= - \frac{4}{3}\frac{\epsilon}{\tau}+\frac{\Phi}{\tau} \label{1st},\\
\partial_\tau \Phi &= - \frac{\Phi}{\tau_\Pi} + \frac{4 \eta}{3 \tau_\Pi \tau}- \frac{4}{3 \tau}\Phi - \frac{\lambda_1}{2\tau_\Pi \eta^2}\Phi^2, \label{2nd}
\end{align}
with longitudinal and transverse pressure $P_L = \frac{1}{3}\epsilon - \Phi$ and $P_T = \frac{1}{3}\epsilon + \frac{1}{2} \Phi$. First order hydrodynamics corresponds to setting $\Phi = \frac{4 \eta}{3\tau}$ in Eq.~(\ref{1st}). At weak coupling, the transport coefficients $\eta$, $\tau_{\Pi}$ and $\lambda_1$ are known \cite{York:2008rr,Arnold:2003zc} leaving zero free parameters to fit.
 besides a time when the hydrodynamics is initialized. 
 We initialize the 1st order hydrodynamics at the latest time we have in our simulation and integrate Eq.~(\ref{1st}) backwards in time. For 2nd order hydrodynamics integrating backwards is highly unstable and 
 we initialize the energy density at some arbitrary earlier time and integrate forwards in time. 


\begin{figure}
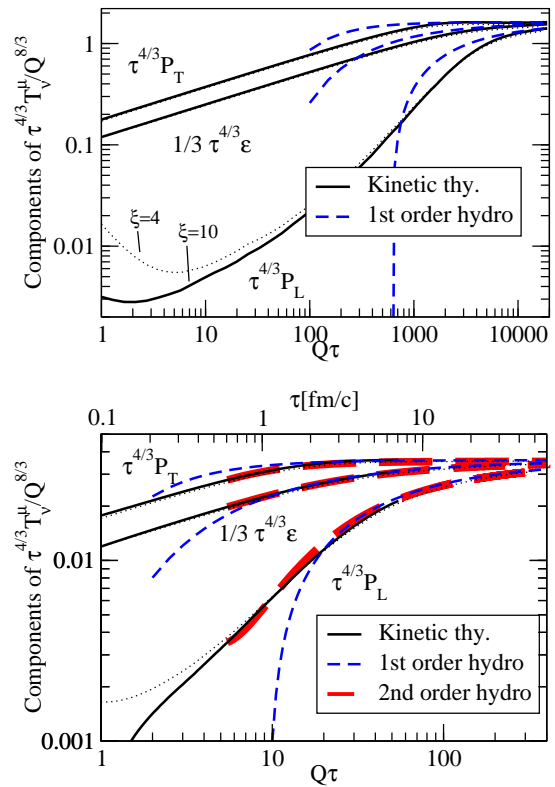

\includegraphics[width=0.4\textwidth]{hydro_F1.eps}\\
\vspace{0.3cm}
\includegraphics[width=0.4\textwidth]{hydro_F10.eps}
\caption{Longitudinal pressure $P_L$, energy density $\epsilon$, and transverse pressure $P_T$ from a simulation with $\xi=10.0$ and $\lambda=1$ (top panel) and $\lambda=10$ (bottom panel). The components of $T^\mu_{\nu}$ have been scaled by $\tau^{4/3}$ so that the ideal hydrodynamics corresponds to horizontal lines. The scale on x-axis with [fm/c] corresponds to $Q_s=2.0$GeV$\approx 10$/fm. The simulations with $\xi=4$ are also displayed with thin dotted lines.}
\label{hydro}
\end{figure}

In Figure \ref{hydro} we examine the validity of the hydrodynamical expansion at small $\lambda=1$ and at realistic intermediate $\lambda = 10$ ($\alpha_s \approx 0.3$) values of coupling. 
In both cases we see that the evolution of the components of the energy momentum tensor asymptotes to their hydrodynamical values. In case of $\lambda=1$, the hydrodynamical behaviour is reached only at a rather late time $Q_s\tau \sim 2000$. We have checked that including 2nd order terms before this time does not make the convergence significantly better; before this time the evolution differs qualitatively from the hydrodynamical prediction. However, rather remarkably, for $\lambda = 10$ even 1st order hydrodynamics gives a very good description of the data all the way to very early times $Q_s\tau \sim 10$ (corresponding to $\tau\sim$1fm/c for $Q_s=2.0$GeV) where the ratio of the pressures is still as large as $P_T/P_L \approx 5$. In addition, including the second order terms significantly improves the convergence. Indeed, we find that initializing the 2nd order hydrodynamics at $Q_s\tau=1$ leads to only 10\% relative error in the energy density at late times.

\section{Discussions and conclusion}

\begin{figure}
\includegraphics[width=0.4\textwidth]{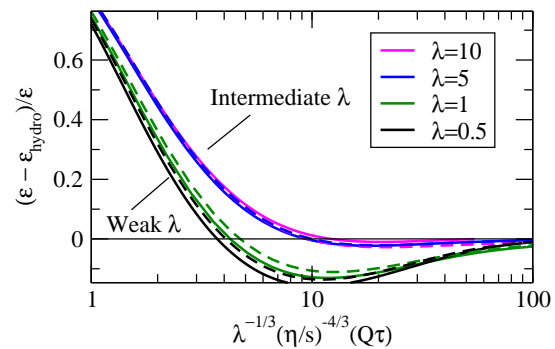}
\caption{
Scaling of the approach to 1st order hydrodynamics at various values of $\lambda$. 
The dashed lines correspond to $\xi =4$ whereas the solid ones have $\xi=10$.
\label{hydro_diff}
}
\end{figure}
The parametric estimate of Baier et al. \cite{Baier:2000sb} for the time when the hydrodynamic behaviour sets is $Q_s\tau \sim \lambda^{-13/5}$. This arises from equating the age of the system $\tau$ with the time scale $\tau_Q$ it takes to affect appreciably the scale $Q_s$ in a thermal bath whose temperature depends on this time $T(\tau) \sim \lambda^{-1/4}Q_s(Q_s\tau)^{-1/4}$ according to conservation of comoving energy density. 
In  \cite{Baier:2000sb}  it was assumed that the rate for affecting the scale $Q_s$ is LPM suppressed \cite{LPM}, giving $\tau_Q \sim (Q_s/T)^{1/2}/\lambda^2 T $. A selfconsistent solution of these equations gives the aforementioned estimate of \cite{Baier:2000sb}.  Arnold and Lenaghan \cite{Arnold:2004ih} noted that as the scattering rate in thermal plasma is $\tau_Q\sim 1/\lambda^2 T$, there can be no process that would make the estimate faster than $Q_s \tau \sim \lambda^{-7/3}$. We have examined the validity of these scaling estimates by plotting the difference of the energy density obtained from the simulation and the 1st order hydrodynamic estimate and find that both of these estimates describe the data poorly. 
However, if we assume that the approach is governed by the hydrodynamical relaxation time  $\tau_Q \sim \tau_{\Pi} \sim \eta/sT$ we get an estimate $\tau \sim \lambda^{1/3}(\eta/s)^{4/3}/Q_s$. Figure \ref{hydro_diff} displays the deviation of the hydrodynamics as a function of rescaled time. In particular for intermediate couplings $\lambda=5,10$ the overlap of the different curves indicates correct scaling. Note that this estimate is parametrically the same as the estimate of Arnold and Lenaghan as parametrically $\eta/s \sim \lambda^{-2}$. However, there are large 
corrections beyond the parametric estimate in $\eta/s$ due to which it is important to use the full numerical result instead of the simpler parametric estimate.
We however believe (in the absence of plasma instabilities) that the correct scaling at sufficiently small $\lambda$ is that of  \cite{Baier:2000sb}. This estimate is however based on a large scale separation of $T(\tau_{hydro})/Q_s \sim \lambda^{2/5}$. Numerically this ratio is not very large even in our simulation with small $\lambda = 1.0$, where the ratio is $T/Q_s \approx 0.15$ (at $Q_s \tau =1800$) as can be inferred from 
Figure \ref{hydro}. Therefore we believe that the scaling predicted by \cite{Baier:2000sb} sets in only at significantly smaller couplings.

In conclusion, we have shown how a far from equilibrium overoccupied configuration of gluons reaches hydrodynamical flow under longitudinal expansion in a weak coupling setting that is systematically improvable. It has been demonstrated by Chesler and Yaffe within AdS/CFT framework that at large values of 't Hooft coupling, hydrodynamics is surprisingly robust even in the presence of large anisotropy \cite{Chesler:2010bi}. The main deliverable of this paper is to show that this robustness is present also in the weak coupling picture extrapolated to intermediate couplings relevant for heavy-ion collisions.

\begin{acknowledgements}
The authors thank Stefan Floerchinger,  Liam Keegan, Tuomas Lappi, Guilherme Milhano, Guy Moore, and Urs Wiedemann for useful discussions. Y.Z. is supported by the European Research Council grant HotLHC ERC-2011-StG-279579.
\end{acknowledgements}

\end{document}